\documentclass[twocolumn,prl,showpacs,amsmath,amssymb,nofootinbib]{revtex4}
\usepackage{amsthm}

\theoremstyle{plain}
\newtheorem{Prop}{Proposition}

\begin{document}
\title{Practical Scheme To Share A Secret Key Through An Up To 27.6\% Bit
 Error Rate Quantum Channel}
\author{H. F. Chau}
\email{hfchau@hkusua.hku.hk}
\affiliation{Department of Physics, University of Hong Kong, Pokfulam Road,
 Hong Kong.}
\date{\today}

\begin{abstract}
 A secret key shared through quantum key distribution between two cooperative
 players is secure against any eavesdropping attack allowed by the laws of
 physics. Yet, such a key can be established only when the quantum channel
 error rate due to eavesdropping or imperfect apparatus is low. Here, I report
 a practical quantum key distribution scheme making use of an adaptive privacy
 amplification procedure with two-way classical communication. Then, I prove
 that the scheme generates a secret key whenever the bit error rate of the
 quantum channel is less than $0.5-0.1\sqrt{5} \approx 27.6\%$, thereby making
 it the most error resistant scheme known to date.
\end{abstract}

\maketitle
 Quantum key distribution (QKD) is the process of sharing a secret bit string,
 known as the key, between two cooperative players, commonly called Alice and
 Bob, by exchanging quantum signals. Since an unknown quantum state cannot be
 perfectly cloned \cite{noclone1,noclone2}, any eavesdropping attempt by Eve
 will almost surely disturb the transmitted quantum states. Thus, by carefully
 estimating the error rate of the transmitted quantum states, Alice and Bob
 know with great confidence the quantum channel error rate, which in turn
 reflects the eavesdropping rate. (In contrast, Alice and Bob can never be sure
 if Eve has eavesdropped in classical key distribution because classical
 signals can be copied without being caught in principle.) If the estimated
 eavesdropping rate is high, they abort the scheme and start over again. On the
 other hand, if the estimated eavesdropping rate is low, privacy amplification
 procedure such as quantum error correction or entanglement purification can be
 used to distill out an almost perfectly secure key
 \cite{lochauqkdsec,biasedbb84,mayersjacm}.
 
 It is instructive to devise a secure QKD scheme that tolerates as high a
 quantum channel error rate as possible and subject that scheme to a vigorous
 cryptanalysis. Indeed, Mayers \cite{mayersjacm} and Biham \emph{et al.}
 \cite{biham} proved the security of the so-called BB84 QKD scheme \cite{bb84}
 against all kinds of attack allowed by the laws of quantum physics. Following
 Mayers' proof, a provably secure key is established whenever the channel error
 rate is less than about 7\%. Lo and Chau proved the security of an
 entanglement-based QKD scheme \cite{lochauqkdsec}. By scrambling the qubits
 before transmission and using the quantum Gilbert-Varshamov argument for a
 general quantum stabilizer code \cite{GF4code}, the Lo and Chau scheme
 tolerates up to about 18.9\% channel error. Nonetheless, the Lo and Chau
 scheme requires quantum computers and hence is not practical at the present
 moment. By beautifully combining the essences of the Mayers as well as Lo and
 Chau proofs, Shor and Preskill gave an ingenious security proof of the BB84
 scheme that applies up to 11.0\% channel error \cite{shorpre}. The most error
 resistant QKD scheme known to date was recently found by Gottesman and Lo.
 Built upon the Shor-Preskill proof, Gottesman and Lo showed that a carefully
 designed privacy amplification procedure with two-way communication increases
 the error tolerant level of a QKD scheme. In particular, they proved that the
 six-state QKD scheme introduced by Bru\ss~\cite{sixstate} tolerates up to
 about 23.7\% bit error rate (or equivalently up to about 35.5\% channel error
 rate) \cite{qkd2waylocc}. Recently, Gottesman and Lo further improved their
 two-way communication protocol and showed that it generates a provably secure
 key up to 26.4\% bit error rate \cite{new}. (Here, the channel error rate and
 bit error rate refer to the rate of quantum and spin flip errors occurring in
 the insecure noisy quantum channel respectively.)

 Here, I report an adaptive privacy amplification procedure for the six-state
 scheme. Then, I prove that this procedure enables the six-state scheme to
 generate a provably secure key up to $0.5-0.1\sqrt{5} \approx 27.6\%$ bit
 error rate (or equivalently up to $0.75-0.15\sqrt{5} \approx 41.4\%$ quantum
 channel error), breaking the 26.4\% bit error rate record of Gottesman and Lo.
 This scheme is also practical, requiring no quantum computer or search for
 asymptotically good quantum codes. Since no BB84-based scheme can tolerate
 more than 25\% bit error rate \cite{new}, the 27.6\% bit error rate tolerable
 six-state scheme reported here convincingly demonstrates the advantage in
 error tolerability of the six-state scheme over BB84. 

 Before reporting the adaptive procedure, let me briefly review the privacy
 amplification procedure introduced by Gottesman and Lo \cite{qkd2waylocc}.
 In the first step of the Gottesman-Lo privacy amplification procedure, Alice
 and Bob perform entanglement purification with local quantum operation and
 two-way classical communication (LOCC2 EP). Specifically, they randomly pair
 up their corresponding bits in the string and compare the result of a
 bilateral exclusive or (BXOR) in each pair. They keep their corresponding
 control bits in each pair only if their parities agree. In the second step,
 Alice and Bob apply the $[3,1,3]_2$ phase error correction (PEC). This is
 equivalent to randomly forming trios of the remaining bits and replace each
 trio by their corresponding parities \cite{qkd2waylocc}. Alice and Bob apply
 LOCC2 EP and PEC alternatively until the error rate of the resultant signal
 can be handled by an asymmetric Calderbank-Shor-Steane (CSS) quantum code
 \cite{CSSqBound,steane} with great confidence. Then, they apply the
 Shor-Preskill error correction procedure \cite{shorpre} to the remaining bits
 using the above CSS code. By doing so, they end up sharing a secret key with
 exponentially close to 100\% confidence. Gottesman and Lo further showed that
 their procedure brings down the error rate whenever the channel error rate is
 less than about 23.7\% \cite{qkd2waylocc}.

 The Gottesman-Lo two-way privacy amplification procedure reviewed above can be
 improved in two ways. First, there is no reason why one must apply LOCC2 EP
 and PEC alternately. Instead, Alice and Bob should devise a suitable privacy
 amplification procedure based on the estimated $\sigma_x$, $\sigma_y$ and
 $\sigma_z$ error rates of the qubits transmitted through the insecure noisy
 channel. Besides, they may use $[r,1,r]_2$ for some $r > 3$ as their phase
 error correction code. In fact, using this approach, Gottesman and Lo proved
 that the six-state scheme can tolerant a bit error rate up to 26.4\%
 \cite{new}. Second, although the asymmetric CSS code used by Gottesman and Lo
 is known to exist using Gilbert-Varshamov type of argument \cite{CSSqBound},
 explicitly finding it may be difficult in general. Fortunately, concatenated
 quantum CSS code is already sufficient in handling the final error correction
 in the privacy amplification procedure. More importantly, various concatenated
 quantum CSS codes and their decoding algorithms are known.

 Before I report my six-state scheme, I first call upon two propositions below
 to study the effects of LOCC2 EP and PEC on the error rates of the signal.

\begin{Prop}
 Suppose Alice sends Bob several qubits through a quantum channel whose
 $\sigma_x$, $\sigma_y$ and $\sigma_z$ error rates due to either noise or
 eavesdropping are $p_x$, $p_y$ and $p_z$ respectively. Let $p_I =
 1-p_x-p_y-p_z$. If the error suffered by each qubit is independent of the
 other, then the error rates of the resultant qubits after going through one
 around of LOCC2 EP are given by
 \begin{equation}
  \left\{ \begin{array}{rcl} p^\mathrm{EP}_I & = & \displaystyle\frac{p_I^2 +
   p_z^2}{(p_I + p_z)^2 + (p_x + p_y)^2} , \\ \\
   p^\mathrm{EP}_x & = & \displaystyle\frac{p_x^2 + p_y^2}{(p_I + p_z)^2 +
   (p_x + p_y)^2} , \\ \\
   p^\mathrm{EP}_y & = & \displaystyle\frac{2p_x p_y}{(p_I + p_z)^2 + (p_x +
   p_y)^2} , \\ \\
   p^\mathrm{EP}_z & = & \displaystyle\frac{2p_I p_z}{(p_I + p_z)^2 + (p_x +
   p_y)^2} .
  \end{array}
  \right. \label{E:LOCC2EP_map_asy}
 \end{equation}
 Furthermore, the error rate in each of the resultant qubit after the LOCC2 EP
 is independent of each other. \label{Prop:locc2ep}
\end{Prop}
\begin{proof}
 Recall that in the LOCC2 EP, Alice and Bob randomly pair up their
 corresponding shares of the qubits and apply BXOR to each pair. During the
 BXOR operation, any $\sigma_x$ error in the control qubit remains unaltered.
 In contrast, the $\sigma_z$ error of the resultant control qubit is inherited
 from both the original control and target qubits \cite{NielsenChuang}. 
 Since Alice and Bob reject the pair if the measurement results of their share
 of target qubit differ, hence the remaining control qubit is error-free if the
 error operator acting on the original control and target qubits equals $I
 \otimes I$ or $\sigma_z \otimes \sigma_z$. Similarly, the remaining control
 qubit suffers $\sigma_x$, $\sigma_y$, and $\sigma_z$ errors if the error
 operator acting on the original control and target qubits equals $\sigma_x
 \otimes \sigma_x$ or $\sigma_y \otimes \sigma_y$, $\sigma_x \otimes \sigma_y$
 or $\sigma_y \otimes \sigma_x$, and $I\otimes \sigma_z$ or $\sigma_z \otimes
 I$ respectively. Since error suffered by each qubit is independent of each
 other, hence Eq.~(\ref{E:LOCC2EP_map_asy}) holds. The independence of
 resultant error rates after the LOCC2 EP procedure follows directly from the
 independence of channel error for the qubits received by Bob.
\end{proof}

 By Proposition~\ref{Prop:locc2ep} and mathematical induction, it is
 straight-forward to check that the error rates of the resultant qubits after
 going through $k$ rounds of LOCC2 EP are given by
\begin{equation}
 \left\{ \begin{array}{rcl}
  p_I^{k\,\mathrm{EP}} & = & [ (p_I + p_z)^{2^k} + (p_I - p_z)^{2^k} ] / 2D ,
  \\
  p_x^{k\,\mathrm{EP}} & = & [ (p_x + p_y)^{2^k} + (p_x - p_y)^{2^k} ] / 2D ,
  \\
  p_y^{k\,\mathrm{EP}} & = & [ (p_x + p_y)^{2^k} - (p_x - p_y)^{2^k} ] / 2D ,
  \\
  p_z^{k\,\mathrm{EP}} & = & [ (p_I + p_z)^{2^k} - (p_I - p_z)^{2^k} ] / 2D ,
 \end{array} \right. \label{E:ktimeslocc2ep}
\end{equation}
 where $D = (p_I + p_x)^{2^k} + (p_x + p_y)^{2^k}$. So whenever $p_I > 1/2$,
 $p_I^{k\,\mathrm{EP}} > 1/2$ and $p_z^{k\,\mathrm{EP}} < 1/2$. Further,
 $p_I^{k\,\mathrm{EP}}, p_z^{k\,\mathrm{EP}}\rightarrow 1/2$ and
 $p_x^{k\,\mathrm{EP}}, p_y^{k\,\mathrm{EP}} \rightarrow 0$ as $k\rightarrow
 \infty$. That is, repeated application of LOCC2 EP reduces $\sigma_x$ and
 $\sigma_y$ errors at the expense of possibly increasing $\sigma_z$ and perhaps
 also the overall error rates.
 
\begin{Prop}
 We use the notations in Proposition~\ref{Prop:locc2ep}. Suppose Alice and Bob
 divide their shared pairs into $n$ sets each containing $r$ shared pairs. And
 then they perform one round of PEC using the $[r,1,r]_2$ majority vote phase
 error correction code. The resultant error rates of the signal after one round
 of PEC satisfy
 \begin{equation}
  \left\{ \begin{array}{rcl} p^\mathrm{PEC}_x + p^\mathrm{PEC}_y & \leq & r
   (p_x + p_y) , \\ \\
   p^\mathrm{PEC}_y + p^\mathrm{PEC}_z & \leq & \left[ 4 (p_I + p_z) (p_x +
   p_y) \right]^{r/2} \\
   & \leq & e^{-2 r (0.5 - p_z - p_y)^2} ,
  \end{array}
  \right. \label{E:PEC_map_asy}
 \end{equation}
 provided that $p_I > 1/2$. Also, the error rate in each of the resultant qubit
 after PEC is independent of each other. \label{Prop:pec}
\end{Prop}
\begin{proof}
 The idea of the proof is the same as that in Proposition~\ref{Prop:locc2ep}.
 Recall that the error syndrome of the $[r,1,r]_2$ phase error correction code
 is given by
 \begin{equation}
  \left[ \begin{array}{ccccc}
  1 & 1 \\
  1 & & 1 \\
  \vdots & & & \ddots \\
  1 & & & & 1
  \end{array} \right] . \label{E:PEC_syn}
 \end{equation}
 So, after measuring this error syndrome, the $\sigma_z$ error stays on the
 control qubit while the $\sigma_x$ error propagates from the control as well
 as all target qubits to the resultant control qubit \cite{NielsenChuang}.
 Therefore upon PEC, the resultant control qubit is spin-flip error-free
 whenever there is an even number of qubits amongst the $r$ of them in the same
 set suffering spin-flip error. Hence, the first inequality in
 Eq.~(\ref{E:PEC_map_asy}) holds. Similarly, the resultant control qubit
 suffers from phase-shift error provided that at least $\left\lceil (r-1)/2
 \right\rceil$ out of the $r$ qubits are suffering from phase-shift error. Such
 a probability of occurrence equals $\sum_{a \geq \left\lceil (r-1)/2
 \right\rceil} \binom{r}{a} (p_y + p_z)^a (p_I + p_x)^{r-a}$. Combining with
 the inequality \cite{Roman}
\begin{equation}
 \sum_{k=0}^{\lambda n} \binom{n}{k} p^k (1-p)^{n-k} \leq \lambda^{-\lambda n}
 (1-\lambda)^{-(1-\lambda)n} p^{\lambda n} (1-p)^{(1-\lambda)n}
 \label{E:inequality}
\end{equation}
 for $0 < \lambda < p$, we conclude that the probability of having a phase
 error is upper bounded by $[4 (p_I + p_x) (p_y + p_z)]^{r/2}$. Thus, the first
 line of the second inequality in Eq.~(\ref{E:PEC_map_asy}) is satisfied. To
 arrive at the second line, one simply considers the Taylor series expansion
 of $\log [1 + (2 p_I + 2 p_x - 1)] + \log [1 + (2 p_y + 2 p_z - 1)]$ and uses
 the observation that all odd power terms in the expansion are canceled.
\end{proof}

 Proposition~\ref{Prop:pec} tells us that if $0.5 - p_z - p_y \gg \sqrt{p_x +
 p_y}$, the phase error can be greatly reduced after one round of PEC by
 choosing $r \approx 0.01 / (p_x + p_y)$. Specifically, with this choice of
 $r$, Eq.~(\ref{E:PEC_map_asy}) implies that $p^\mathrm{PEC}_y +
 p^\mathrm{PEC}_z$ is exponentially small while $p^\mathrm{PEC}_x +
 p^\mathrm{PEC}_y$ is at most about 1\%.
 
 Alice and Bob may exploit the dynamics of LOCC2 EP and PEC to perform their
 privacy amplification. Specifically, they first repeatedly apply LOCC2 EP
 until $0.5 - p_z - p_y \gg \sqrt{p_x + p_y}$. Then, applying PEC once will
 bring the overall error rate $p_x + p_y + p_z$ down to an acceptable value.
 And then, Alice and Bob may choose to use the concatenated Steane's seven
 qubit code in the Shor-Preskill procedure. Recall that Steane's seven qubit
 code corrects one error out of seven qubits \cite{steane}. Thus, as long as
 Alice and Bob randomly permute the bits before applying the Shor-Preskill
 procedure, the overall error rate that is almost surely tolerated by the
 concatenated Steane's seven qubit code is equal to the smallest positive root
 of the equation
\begin{equation}
 1-\lambda = (1-\lambda)^7 + 7 (1-\lambda)^6 \lambda , \label{E:steaneerror}
\end{equation}
 namely, about 5.8\%. The upshot is that the error correction algorithm for the
 concatenated Steane's seven qubit code is known and can be carried out
 efficiently.

 With these two improvements in mind, I write down my modified six-state scheme
 below.

\begin{enumerate}
 \item Alice prepares $N$ qubits each randomly chosen from $|0\rangle$,
  $|1\rangle$, $|0\rangle\pm|1\rangle$ and $|0\rangle\pm i|1\rangle$ and sends
  them to Bob \cite{sixstate}. Bob acknowledges the reception of the qubits and
  measures each of them randomly and independently along one of the following
  three bases: $\{ |0\rangle, |1\rangle$\}, $\{ |0\rangle\pm |1\rangle \}$ and
  $\{ |0\rangle\pm i|1\rangle \}$. Then, Alice and Bob publicly announce the
  bases they have used to prepare or measure each qubit. They keep only those
  qubits that are prepared and measured in the same basis.
  \label{Scheme:Prepare}
 \item Alice and Bob estimate the channel error rate by sacrificing a few
  qubits. Specifically, they divide the qubits into three sets according to
  their bases of measurement. They randomly pick $\mbox{O} (\log [1/\epsilon])$
  qubits from each set and publicly compare the preparation and measurement
  results of each chosen qubit. In this way, they know the estimated channel
  error rate with standard deviation $\epsilon$. (Detail proof of this claim
  can be found in Ref.~\cite{biasedbb84}.) If the estimated channel error rate
  is too high, they abort the scheme and start over again. \label{Scheme:QC}
 \item Using the convention that $|0\rangle$, $|0\rangle - |1\rangle$ and
  $|0\rangle - i |1\rangle$ represent a logical 0 while the $|1\rangle$,
  $|0\rangle + |1\rangle$ and $|0\rangle + i |1\rangle$ represent a logical 1,
  Alice and Bob convert their untested measured qubits into secret strings.
  Then, they perform the following privacy amplification procedure on their
  secret bit strings.
  \begin{enumerate}
   \item They apply the LOCC2 EP procedure proposed by Gottesman and Lo in
    Ref.~\cite{qkd2waylocc}. Specifically, they randomly pair up their
    corresponding secret bits and announce the parities of each pair. They keep
    the control bit in each pair only if their announced parities for the pair
    agree. They repeat the above LOCC2 EP procedure until there is an integer
    $r > 0$ such that the estimated quantum channel error given by
    Eq.~(\ref{E:PEC_map_asy}) is less than 5\%. They abort the scheme either
    when such an integer $r$ is greater than the number of remaining bits they
    have or when they have used up all their bits in this procedure.
    \label{step:ep}
   \item They apply the PEC procedure introduced by Gottesman and Lo in
    Ref.~\cite{qkd2waylocc} using the $[r,1,r]_2$ majority vote phase error
    correction code once. Specifically, Alice and Bob randomly divide the
    resultant bits into sets each containing $r$ bits. They replace each set by
    the parity of the $r$ bits in the set. \label{step:pec}
   \item Alice and Bob randomly permute the order of their remaining bits and
    apply the Shor-Preskill privacy amplification procedure \cite{shorpre} to
    these bits with the concatenated Steane's seven qubit code. The level of
    concatenation depends on the estimated worst case $p_x + p_y + p_z$ given
    by Eq.~(\ref{E:PEC_map_asy}) and the final required fidelity of the state.
    Specifically, suppose that the concatenated Steane's seven qubit code is
    constructed from two binary classical codes $C_1$ and $C_2$ satisfying $C_2
    \subset C_1$. Alice randomly picks a codeword $u \in C_1$ and publicly
    announces the sum of $u$ and her remaining bit string modulo 2. Bob
    subtracts Alice's announced bit string from his own remaining bit string
    modulo 2; and then he applies the $C_2$ error correction to recover the
    codeword $u \in C_1$. They use the coset $u+C_2$ as their secret key.
    \label{step:qecc}
   \end{enumerate}
  \label{Scheme:PA}
\end{enumerate}

 To prove the security of the above scheme, I follow the arguments of
 Refs.~\cite{lochauqkdsec,shorpre,sixstateproof,qkd2waylocc}. First, since this
 is a prepare-and-then-measure scheme, any Eve's quantum cheating strategy can
 be reduced to a classical one \cite{lochauqkdsec,sixstateproof}. Second, Eve
 does not know how Alice and Bob group the qubit pairs in LOCC2 EP and PEC
 beforehand. Hence, the resultant error rate after going through either LOCC2
 EP or PEC depends only on the probabilities of $\sigma_x$, $\sigma_y$ and
 $\sigma_z$ errors and the number of qubits transmitted
 \cite{lochauqkdsec,qkd2waylocc}. Thus, to study the asymptotic error tolerable
 rate of the above scheme, it suffices to consider cheating strategies
 characterized only by $p_x$, $p_y$ and $p_z$ respectively. Since Alice chooses
 the six states randomly and uniformly, the untested qubits can be regarded as
 having passed through a depolarizing channel \cite{qkd2waylocc}. Hence, Alice
 and Bob almost surely know that $p_x = p_y = p_z$ for their untested qubits.

 From Eq.~(\ref{E:PEC_map_asy}) in Proposition~\ref{Prop:pec}, I know that
 after applying LOCC2 EP k times, PEC will bring the quantum error rate down
 to, say, 5\% if $r = 0.04 / (p_x^{k\,\mathrm{EP}} + p_y^{k\,\mathrm{EP}})$ and
 $2 r (0.5 - p_z^{k\,\mathrm{EP}} - p_y^{k\,\mathrm{EP}}) \gg 1$. Putting $p_x
 = p_y = p_z = (1-p_I)/3$ into Eq.~(\ref{E:ktimeslocc2ep}), I conclude that
 this is possible when $k \rightarrow\infty$ and $(p_I - p_z)^2 > (p_I + p_z)
 (p_x + p_y)$. This condition implies that $20 p_I^2 - 10 p_I - 1 > 0$ or $p_I
 > 0.25 + 0.15\sqrt{5}$. In other words, the above scheme tolerates a bit error
 rate up to $p_x + p_y = 0.5 - 0.1\sqrt{5} \approx 27.6\%$ (which corresponds
 to a quantum channel error rate of $p_x + p_y + p_z = 0.75-0.15\sqrt{5}
 \approx 41.4\%$).

 Besides, once Alice and Bob estimate the channel error rates, then they can
 efficiently compute the number of LOCC2 EP to be applied as well as the level
 of concatenation for the Steane's seven qubit code to be used. Finally, the
 error syndrome of the concatenated Steane's seven qubit code as well as the
 corresponding Shor-Preskill procedure are straight-forward to compute.

 The 27.6\% bit error rate bound reported here shows that the six-state scheme
 is more noise resistant than the BB84 scheme since no BB84 scheme can tolerate
 more than 25\% bit error \cite{new}. In addition, the adaptive privacy
 amplification idea can be applied to increase the error tolerant level in a
 number of QKD schemes. For instance, the above adaptive privacy amplification
 procedure enables the BB84 to generate a provably secure key whenever the bit
 error rate is less than 20.0\% (or equivalently, a quantum channel error rate
 of less than 39.9\%). Besides, one can show the existence of a biased
 entanglement-based QKD scheme requiring quantum computers whose key is
 provably secure whenever the bit error rate is less than 33.3\%
 \cite{further}.

\begin{acknowledgments}
 This work is supported in part by the HKU Outstanding Young Researcher Award
 and the RGC grant HKU~7095/97P of the HKSAR government. I would like to thank
 H.-K. Lo for explaining his work with Gottesman on applying two-way classical
 communication to quantum key distribution and for telling me his recent work
 with Gottesman \cite{new} prior to its public dissemination, D. Gottesman and
 H.-K. Lo for their valuable discussions on privacy amplification, D.~W.~C.
 Leung and H.-K. Lo for their critical reading of this manuscript.
\end{acknowledgments}

\bibliography{qc29.5}
\end{document}